\documentclass[preprint,12pt,groupaddress]{elsarticle}

\usepackage{booktabs}
\usepackage{color}
\usepackage{lipsum}
\usepackage{float}
\usepackage[titletoc,toc,title]{appendix}
\usepackage{amssymb}
\usepackage{epsfig}
\usepackage{amsmath}
\usepackage{times}
\usepackage{graphicx}
\usepackage{booktabs}
\usepackage{colortbl}
\usepackage{natbib}

\begin{document}

\title{Nonlinear cancer chemotherapy: modelling the Norton-Simon hypothesis}

\author[label1]{\'{A}lvaro G. L\'{o}pez}
\author[label2,label21]{Kelly C. Iarosz}
\author[label3]{Antonio M. Batista}
\author[label1]{Jes\'{u}s M. Seoane}
\author[label4]{Ricardo L. Viana}
\author[label1,label5]{Miguel A. F. Sanju\'an}

\address[label1]{Nonlinear Dynamics, Chaos and Complex Systems Group.\\Department of Physics, Universidad Rey Juan Carlos, Tulip\'an s/n, 28933 M\'ostoles, Madrid, Spain}

\address[label2]{Physics Institute, University of S\~ao Paulo, 05508-090 S\~ao Paulo, SP, Brazil}
\address[label21]{Department of Physics, Humboldt University, 12489 Berlin, Germany}

\address[label3]{Department of Mathematical and Statistics, State University of Ponta Grossa, 84030-900 Ponta Grossa, PR, Brazil}

\address[label4]{Department of Physics, Federal University of Paran\'{a}, 81531-990 Curitiba, PR, Brazil}

\address[label5]{Institute for Physical Science and Technology, University of Maryland, College Park, Maryland 20742, USA}

\date{\today}

\begin{abstract}
A fundamental model of tumor growth in the presence of cytotoxic chemotherapeutic agents is formulated. The model allows to study the role of the Norton-Simon hypothesis in the context of dose-dense chemotherapy. Dose-dense protocols aim at reducing the period between courses of chemotherapy from three weeks to two weeks, in order to avoid tumor regrowth between cycles. We address the conditions under which these protocols might be more or less beneficial in comparison to less dense settings, depending on the sensitivity of the tumor cells to the cytotoxic drugs. The effects of varying other parameters of the protocol, as for example the duration of each continuous drug infusion, are also inspected. We believe that the present model might serve as a foundation for the development of more sophisticated models for cancer chemotherapy.\\
\end{abstract}

\maketitle



\section{Introduction}\label{sec:intro}

Improvements in chemotherapy mainly depend upon minimizing the toxic side-effects of drugs on healthy tissues and tackling the resistance of cancer cells to such medications \cite{lipo,drr}. Nevertheless, randomized trials carried out along the last half century have proved that the specific arrangement of protocols is of relevance as well \cite{dd1,dd2,dd3}. In particular, breast cancer trials have revealed that reducing the period between cycles of chemotherapy from three to two weeks introduce moderate but statistically significant benefits in disease free survival at five years and overall survival \cite{dd4}. Surprisingly, this can be achieved without introducing higher toxicities with the aid of granulocyte colony-stimulating factor. This increase in the frequency of drug administration has been termed \emph{dose-dense chemotherapy}, because when the dose is represented against time, these protocols look more dense \cite{norton}.

Despite recent success in adjuvant chemotherapy for breast cancer \cite{clinclast}, the benefits of dose-dense chemotherapy are not undoubted \cite{notwo}. Remarkably, the appearance of these protocols for the treatment of solid tumors has posed fundamental questions concerning the nature of solid tumor growth. Moreover, it has also suggested a theory of tumor growth for breast cancer, based on the concept of self-metastasis \cite{norton,ender}. In particular, the \emph{Norton-Simon hypothesis} has been suggested as a possible explanation to its beneficial properties. This hypothesis states that the rate of destruction by chemotherapy is proportional to the rate of growth of the unperturbed tumor \cite{nsh}. As has been recently suggested, the concept of dose-density might be pertinent even in the absence of any hypothesis affecting the particular nature of the growth of the tumor \cite{ddpc}.

Concerning previous modelling efforts on this topic, some preliminary discussion is deserved. Several deterministic and stochastic models have appeared in the literature over the last decades to describe different phenomena related to dose-dense chemotherapy and protocol optimization \cite{train,castor,monro,dono,bft,cold}. However, some of these works do not model cell kill \cite{train}, while others model it in an indirect way, by imposing conditions on the decay of the tumor once the drugs have been administered \cite{castor}. On the other hand, to the best of our knowledge, those works that explicitly model cell kill and consider the Norton-Simon hypothesis always assume a linear dependence between the rate of cell destruction and the dose of drug delivered \cite{monro,dono}. This assumption leads to dose-response curves which are linear when plotted on a log scale. These curves, as has been demonstrated, do not fit properly data from solid tumors \cite{gard}.

In the present work we propose an ordinary differential equation (ODE) model for nonlinear cancer chemotherapy to investigate under which conditions dose-dense protocols are more or less advantageous. It is not our purpose to develop specific protocols for a certain class of tumors, but rather to find general guiding principles which might aid clinicians to bias their decisions in the design of better protocols in the future. Special attention is paid to the relative possible benefits of dose-dense protocols compared to more dose-intense ones. Although dose-intensity is commonly defined as an average over the whole treatment in the field of oncology, we use it hereafter to deem an increase of the total dose administered per cycle \cite{ddpc}.

\section{Model description}\label{sec:md}

\subsection{Tumor growth in the presence of chemotherapeutic drugs}\label{sec:md}

A general multicompartment model for cancer chemotherapy can be built from previous modeling works \cite{panad, pinho, gard2,kav,kavg,dono}. Since the cell cycle comprises four phases (mitosis, gap 1, synthesis and gap 2) and cells can also be quiescent (gap 0), a possible modeling framework should consider five different compartments representing the cell populations in each of them. In principle, the use of different compartments is relevant since some drugs only destroy cells on a specific phase of the cell cycle \cite{mabio,polans}. However, in practice, one or more compartments can be joined into a single compartment, reducing the dimensionality of the problem. For example, if there is only one cycle specific drug, a model with three compartments can be devised. Firstly, two compartments would be required to distinguish between mitotic and quiescent cells. Then, the cycling cells should be subdivided among those cells which are in the particular phase in which the cycle specific drugs exert their effect, and those that are in the complementary part of the cell cycle.

We begin with the most simple model example capable of representing the Norton-Simon hypothesis. Thus, in this first approach, we are only interested in the effects of just one cycle non-specific (CNS) drug and, therefore, we restrict ourselves to a single compartment of mitotic tumor cells $P(t)$. Accordingly, we also gather in a single compartment the mitotic and the quiescent cells, and use a single function $\gamma(P)$ to represent the net result of proliferation and death of tumor cells due to necrosis and apoptosis. The drug concentration is represented as $C(t)$. Therefore, we study the differential equation
\begin{equation}
\dfrac{d P}{d t}= \gamma(P) P -\kappa(P,C) P.
\label{eq:2}
\end{equation}
Because a somatic cell divides through mitosis, the growth of a cell population is at most geometric \cite{tumrinm}. This suggests that $\gamma(P)$ should be a monotonically decreasing function of $P$. We illustrate most of our analysis using $\gamma(P)=r (1-P/K)$, which corresponds to a logistic growth \cite{onlaw,ramcon,bcd}, even though Gompertzian growth $\gamma(P)=r \log(K/P)$ is tested as well \cite{gomp}. 

The design of the function $\kappa(P,C)$ is more delicate, since we want the model to be capable of reproducing the Norton-Simon hypothesis. We start from the Exponential-Kill model, which has been tested using \emph{in vitro} and \emph{in vivo} data \cite{gard,validbul}. The fractional cell kill is represented in such model by a term $b (1-e^{-\rho C})$, where $b$ represents the maximum fractional cell kill and $\rho$ is commonly referred in the literature as the resistance of the tumor cells to the drugs \cite{ddpc,gard}. However, resistance is a complex evolutionary process involving several biological mechanisms, as for example the drug efflux through transmembrane proteins or DNA-damage repair mechanisms \cite{drca}. Hence, those models that incorporate tumor resistance to chemotherapeutic drugs traditionally make use of several compartments \cite{castor, monro} consisting of cells with different values of $\rho$. Therefore, we shall call the parameter $\rho$ the sensitivity, hereafter. More precisely, $\rho$ is directly proportional to the sensitivity. In the present work we investigate rather homogeneous tumors consisting of cells whose deviations in this parameter from the average value is very small. In summary, we can consider a nonlinear separable function in the form $\kappa(P,C)=h(P)b(1-e^{-\rho C})$. It is important to note that this new fractional cell kill saturates for increasing values of the drug concentration, a feature  frequently overlooked in models of chemotherapy for solid tumors, which mainly consider a linear dependence on the drug concentration \cite{monro,dono}.

Now we recall that, according to the Norton-Simon hypothesis, the rate of destruction of the tumor is proportional to the rate of growth of the the same unperturbed tumor. Here, we take this statement in a non-strict sense, since we have not found experimental evidence of a direct proportionality $h(P) \propto \gamma(P)$. Quite the opposite, there is evidence that CNS drugs act to some extent on the quiescent compartment as well \cite{gard2,prp}, which contradicts such direct relation of proportionality. Nevertheless, it is likely that cells inside a solid tumor are less susceptible to chemotherapeutic drugs. This might be a consequence of the fact that the concentration of drugs attenuates as they penetrate inside a solid tumor, since they are absorbed by the outmost external cells. Note also that cells in the interior are less exposed as well and that the cells on the boundary of a tumor are more actively proliferating than the cells in its core \cite{bru,bru2}. Hence, as previously stated, we believe that it is enough to assume that the fractional cell kill of chemotherapeutic drugs decreases for bigger tumors. 

In conclusion, we assume that $h(P)$ is a monotonically decreasing function of the cell population $P$. More specifically, we consider a Holling type II functional response \cite{hol} in the form
\begin{equation}
h(P)=\dfrac{r K}{K+s P}.
\label{eq:3}
\end{equation}
This particular choice can be justified on the following grounds:

\begin{enumerate}[(a)]

\item The maximum rate of cell kill is proportional to the maximum growth rate of the tumor.

\item The fractional cell kill is smaller as the tumor gets close
to the carrying capacity. The maximum value at the carrying
capacity is $r/(1+s)$. Therefore, the parameter $s$ controls the Norton-Simon hypothesis, being both in a direct relation.

\item For $s=0$ or $K \rightarrow \infty$ we obtain the fractional cell kill used in \cite{gard}. In fact, if both conditions hold, the model itself is the one used in such reference. This situation corresponds to an exponentially growing tumor for which the Norton-Simon hypothesis establishes that the rate of destruction should be exponential.

\item The action of some cytotoxic drugs resembles enzyme kinetics, where the drug plays the role of the enzyme and the cell resembles to the substrate (\emph{e.g.} the alkylating agent Cisplatin binds to the cell DNA causing intrastrand cross-links that ultimately can lead to apoptosis). 

\item In the limit of small concentrations or low sensitivity, the term $1-e^{-\rho C}$ can be approximated as $\rho C$. In this limit the function $\kappa(P,C)$ obeys the Michaelis-Menten kinetics, which has been extensively used in previous studies \cite{pinho,kav}.

\end{enumerate}

Nevertheless, other functional responses will be tested, which allow to represent the Norton-Simon effect in its more strict formulation. But, unless otherwise specified, we shall utilize the nonlinear ODE model
\begin{equation}
\dfrac{d P}{d t}=r \left(1-\dfrac{P}{K} \right) P -r b \dfrac{K}{K+s P}\left(1-e^{-\rho C}\right) P,
\label{eq:4}
\end{equation}
throughout the following sections.

\subsection{Pharmacokinetics and protocols of chemotherapy}

It remains to be introduced the pharmacokinetics of the model, which governs the dynamics of the concentration of drug at the tumor site $C(t)$. As in previous works \cite{validbul,mixed,pinho}, we consider a one-compartment model and first order pharmacokinetics, which in some situations can be used as a good approximation in the clinical practice \cite{pks}. The differential equation governing the concentration of the drug is
\begin{equation}
\dfrac{d C}{d t} = I(t) - kC(t),
\label{eq:5}
\end{equation}
where $I(t)$ is the function representing the rate of flow of drug into the body (the instantaneous dose-intensity) and $k$ is the rate of elimination of the drug from the bloodstream, from which the half-life can be computed as $(\log_{e} 2)/k$.
\begin{figure}
\centering
  \includegraphics[width=0.48\linewidth,height=0.25\linewidth]{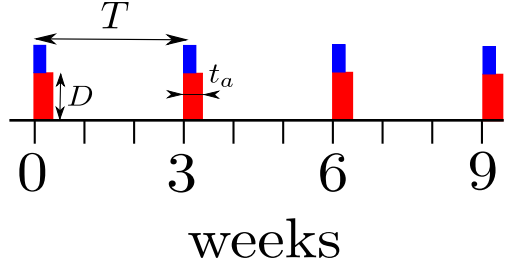}
\caption{\textbf{A protocol of chemotherapy}. The continuous infusion of two drugs (red and blue) delivered with a period $T$ of three weeks and using different doses $D$. The time duration of each infusion $t_{a}$ differs for the two drugs as well.}
\label{fig:1}
\end{figure}

Concerning the drug delivery, we assume that the drug is administered intravenously at constant speed during a time $t_{a}$ in cycles of period $T$. Thus, a typical protocol of chemotherapy can be schematically represented as in Fig.~\ref{fig:1}. If the dose of drug given in a course of chemotherapy is $D$, then the instantaneous dose-intensity $I(t)$ is expressed mathematically as
\begin{equation}
I(t)=\left\{
\begin{aligned}
\dfrac{D}{t_{a}} &~~\text{for}~~t(\text{mod}~T) \in [0,t_{a}) \\
0~&~~\text{for}~~t(\text{mod}~T) \in [t_{a},T)
\end{aligned}.
\right.
\label{eq:6}
\end{equation}

\begin{figure}
\centering
  \includegraphics[width=0.62\linewidth,height=0.33\linewidth]{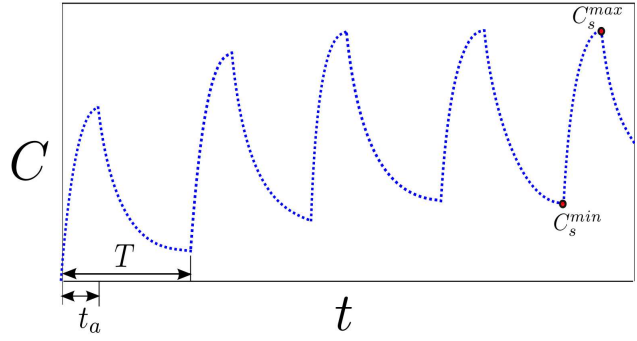}
\caption{\textbf{The drug concentration}. The time series of the drug concentration for five cycles of chemotherapy administered every $T$ weeks. Each course of chemotherapy consists of a continuous infusion of drug that lasts a time $t_{a}$. Hence the rate of infusion is $D/t_{a}$. The constant rate of elimination of the drug from the bloodstream is $k$. Asymptotically, a steady oscillation between two values of the concentration $C_{s}^{\text{min}}$ and $C_{s}^{\text{max}}$ is attained.}
\label{fig:2}
\end{figure}

The solution to equation \eqref{eq:5} with a drug input given by equation \eqref{eq:6}, when a number of $N$ chemotherapeutic cycles have been delivered, yields
\begin{equation}
C(t)=\left\{
\begin{aligned}
\dfrac{D}{k t_{a}}(1-e^{-k t})+a_{N} (t)~~~~~~&~~\text{for}~~t(\text{mod}~T) \in [0,t_{a}) \\
\dfrac{D}{k t_{a}}(e^{k t_{a}}-1)e^{-k t}+a_{N} (t)&~~\text{for}~~t(\text{mod}~T) \in [t_{a},T)
\end{aligned}
\right.,
\label{eq:7}
\end{equation}
where $a_{N}(t)$ represents the decaying accumulated concentration of drug during the $N+1$-th cycle of chemotherapy. This function equals zero during the first cycle ($a_{0}(t)=0$) and it is equal to
\begin{equation}
a_{N} (t)=\dfrac{D}{k t_{a}}(e^{k t_{a}}-1)e^{-k t} \sum_{n=1}^{N}e^{-n k T},
\label{eq:8}
\end{equation}
for the subsequent cycles. Using this equation, it can be easily demonstrated that when the number of cycles $N$ tends to infinity, the concentration reaches a steady oscillation $C_{s} (t)$ between a minimum value
\begin{equation}
C_{s}^{min}=\dfrac{D}{k t_{a}}\dfrac{e^{k t_{a}}-1}{e^{k T}-1},
\label{eq:9}
\end{equation}
and a maximum value $C_{s}^{max}=C_{s}^{min}e^{k(T-t_{a})}$. This periodicity is represented in Fig.~\ref{fig:2}. Since the periodicity of the cycles of chemotherapy commonly spans several weeks, whereas the drug is eliminated rather quickly (days) in comparison, the accumulation of drug is not usually very important, unless $t_{a}$ is close to $T$.
\begin{figure}
\centering
  \includegraphics[width=0.4\linewidth,height=0.38\linewidth]{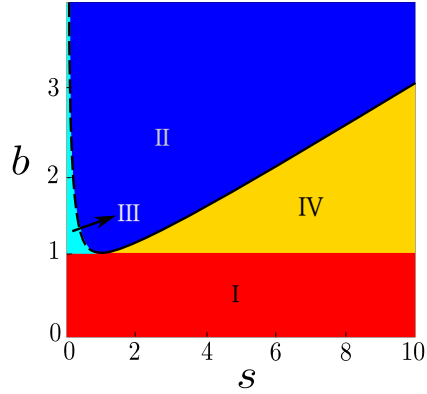}
\caption{\textbf{The parameter set}. When the drugs are given at high doses for long periods of time, four different regions can be distinguished according to the asymptotic dynamics. In region I the therapy is not effective and the tumor is just reduced in size, but not eliminated. The regions II and III represent parameter values for which the tumor is surely destroyed. The cyan region is separated from the dark blue region by the dashed line, because it presents two more fixed points. However, they occur for negative values of the cell population, which have no biological meaning. In region IV the tumor can be eradicated or not depending on its size at the beginning of the treatment.}
\label{fig:3}
\end{figure}

\section{High continuous doses} \label{sec:hcd}

To prepare our intuition for the numerical results that are presented in the following section, we first study the parameter space in the case that of tumors which are highly sensitive ($\rho \rightarrow \infty$) to the drug, which is continuously administered at sufficiently high doses ($D \nrightarrow 0$). If this holds, we can take the limit $\rho C \rightarrow \infty$ and the differential equation \eqref{eq:4} can be simplified to
\begin{equation}
\dfrac{d x}{d \tau}=\left(1-x \right) x - \dfrac{b}{1+s x} x,
\label{eq:2}
\end{equation}
where the dimensionless variables $x=P/K$ and $\tau=r t$ have been introduced. This differential equation possess one, two or three fixed points, depending on the values of the parameters $b$ and $s$. The fixed point $x^{*}=0$ is always present, while the others are the solution to the quadratic equation $(1-x)(1+s x)-b=0$, whose discriminant equals to $(s-1)^2-4 s (b-1)$. As shown in Fig.~\ref{fig:3}, $b<1$ defines region I, in which there always exist two fixed points. The origin is a repelling fixed point, whereas the remaining point is an attractor representing a tumor that exists bellow its carrying capacity. Consequently, if the drug is not effective enough, the tumor can not be destroyed, even if the drugs are delivered continuously and forever. The remaining part of the parameter space is subdivided into three smaller regions (II, III and IV) by the curve $b=1+(s-1)^2/(4 s)$. The upper and left blue regions delimited by this curve (regions II and III) represent a domain for which there is only one positive attracting fixed point $x^{*}=0$. In this case, the tumor can be asymptotically destroyed by the chemotherapeutic drugs. The only difference between regions II and III is that the latter has two more fixed points. However, they occur for negative values of the cell population, which are biologically meaningless. Finally, when we go from region III to region IV, two fixed points are born through a saddle-node bifurcation. Two of them are attractors, one at $x^{*}=0$ and another existing bellow the carrying capacity, separated by a repelling fixed point. Therefore, the tumor can be destroyed or not, depending on its size at the beginning of the treatment.
\begin{figure}
\centering
  \includegraphics[width=0.65\linewidth,height=0.4\linewidth]{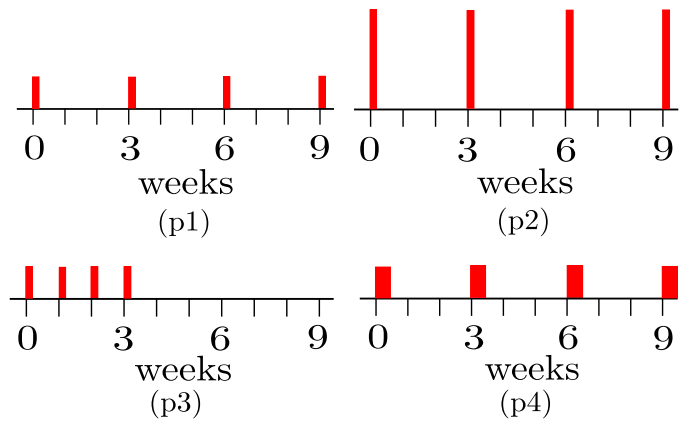}
\caption{\textbf{Investigated protocols}. (p$_{1}$) A reference protocol consisting on a continuous infusion of a dose of a single drug delivered every three weeks through intravenous bolus. (p$_{2}$) A more dose-intense protocol for which the dose of drug per course is multiplied by three. (p$_{3}$) A more dose-dense protocol for which the period between cycles is divided by three. (p$_{4}$) A protocol for which the same dose of drug is delivered, but through longer lasting infusions.}
\label{fig:4}
\end{figure}

\section{Numerical simulations}\label{sec:nsi}

In the present section we accomplish a comparative study between several types of chemotherapeutic protocols. For this purpose, we solve equation \eqref{eq:4} with a drug concentration given by equation \eqref{eq:7} for four different protocols, which are represented in Fig.~\ref{fig:4}, and two sensitivity scenarios. The high sensitivity scenario corresponds to a value $\rho=1.0~\text{mg}^{-1}$, while the low sensitivity scenario is given by $\rho=0.01~\text{mg}^{-1}$, which are within extreme values appearing in \cite{gard}. We use as a reference a protocol (p$_{1}$) with parameter values $T=3~\text{weeks}$, $t_{a}= 1~\text{min}$ and $D=60~\text{mg}$, which are typical for locally advanced breast cancer\footnote{Information about standard protocols of chemotherapy has been drawn from http://www.bccancer.bc.ca/health-professionals/clinical-resources/chemotherapy-protocols}. A second protocol (p$_{2}$) is considered that it is three times more dose-intense, with $D=180~\text{mg}$. This increase in the dose is chosen bearing in mind that the dependence of the fractional cell kill on the drug concentration is exponential. Therefore, this variation suffices for our purposes. Nevertheless, we note that in practice there is a great variability concerning the values of $\rho$ and $D$ among the different drugs and protocols \cite{gard}. The third protocol (p$_{3}$) is more dose-dense than the first protocol, with $T=1~\text{week}$. Although protocols that are more dose-dense that two weeks are unrealistic, we have decided to push dose-density to the smallest values as possible, to have as a reference a situation which is limiting in the time scale commonly considered (weeks). A final fourth protocol (p$_{4}$) tests the effect of increasing the duration of the infusion up to half a week $t_{a}=0.5~\text{weeks}$, keeping the total dose per course constant. 
\begin{figure}
\centering
  \includegraphics[width=0.90\linewidth,height=0.5\linewidth]{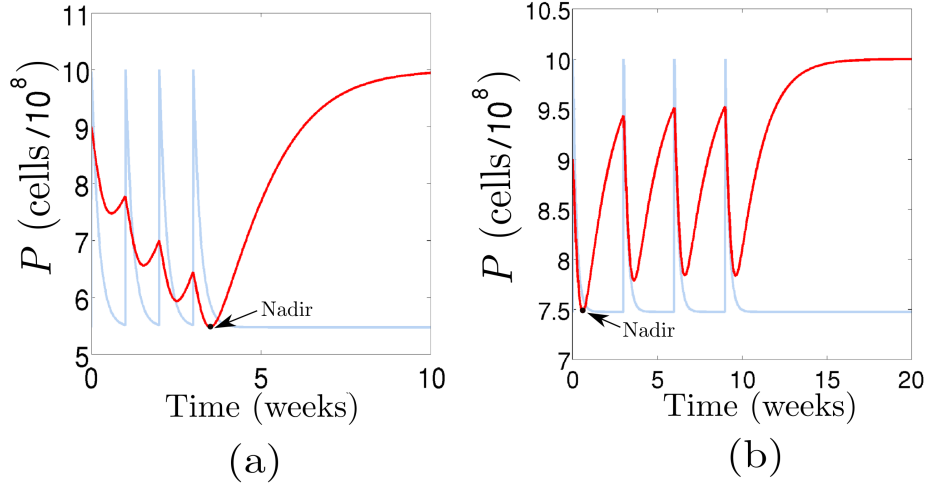}
\caption{\textbf{Time series}. The evolution of a tumor cell population (red) under the action of chemotherapy (blue) for a dose of drug $D=60~\text{mg}$ injected as a bolus ($t_{a}= 1~\text{min}$). The parameter values characterizing the cytotoxicity are $\rho=0.1~\text{mg}^{-1}$, $s=4.1$ and $b=3.6~\text{weeks}^{-1}$. The concentration of drug is plotted for clarity, disregarding its specific values. (a) A dense protocol with a period $T$ of one week. The survival fraction of tumor cells at the minimum of the treatment (nadir) is $\sigma=P(t_{\text{nad}})/P(0)=0.61$.  (b) A less dense protocol with a period $T$ of three weeks. The survival fraction of tumor cells at the nadir is $\sigma=0.84$. Note that in this case the nadir is reached at the beginning of the treatment, while in the previous case the tumor is progressively reduced.}
\label{fig:5}
\end{figure}

To investigate the relative benefits of each protocol, we compute the survival fraction of tumor cells at the nadir $\sigma=P_{\text{nad}}/P_{0}$ of the treatment. This effectiveness criterion has been adopted in previous works \cite{gard2}. Consequently, we recall that, since in this first study we are not explicitly modelling toxicity, the benefits or the effectiveness of a protocol should not be read from a clinical point of view, but instead from a theoretical one. Then, we represent it in the parameter space $(b,s)$, as in the previous section. More specifically, we represent in logarithmic scale the ratio between survival fractions at the nadir $\sigma_{i}/\sigma_{j}$ for a pair ($\text{p}_{i}$ and $\text{p}_{j}$) of protocols. Therefore, when protocol $\text{p}_{i}$ is more beneficial than protocol $\text{p}_{j}$ we have $\log(\sigma_{i}/\sigma_{j})<0$, since $\sigma$ is always smaller than one. Oppositely, it will be higher or equal than zero when $\text{p}_{i}$ is less beneficial than protocol $\text{p}_{j}$.

Unless otherwise specified, the simulations are carried out considering a tumor whose maximum rate of growth is $r=0.8~\text{weeks}^{-1}$ and with a value of the carrying capacity of $K=10^{9}$ cells \cite{validbul}. An extremely fast growing tumor would be one in which all cells were 
ceaselessly dividing through mitosis. Since the cell cycle of a human cell lasts approximately one day, this corresponds to an exponentially growing tumor with a constant rate value of $r=4.85~\text{weeks}^{-1}$. Therefore, we are considering a quite aggressive tumor. The carrying capacity corresponds to a detectable tumor mass of approximately one gram. The original size of the tumor is considered to be $P_{0}=9 \cdot 10^{8}$ cells. We take a typical value of $k=4.85~\text{weeks}^{-1}$, which corresponds to a half-life of approximately one day \cite{ddpc}. Two typical trajectories for the parameter values $\rho=0.1~\text{mg}^{-1}$, $s=4.1$ and $b=3.6~\text{weeks}^{-1}$ are shown in Fig.~\ref{fig:5}. They illustrate a situation suggesting the possible benefits of dose-dense protocols.

\subsection{High sensitivity scenario}\label{sec:lrs}

\begin{figure}
\centering
\includegraphics[width=0.75\linewidth,height=1.0\linewidth]{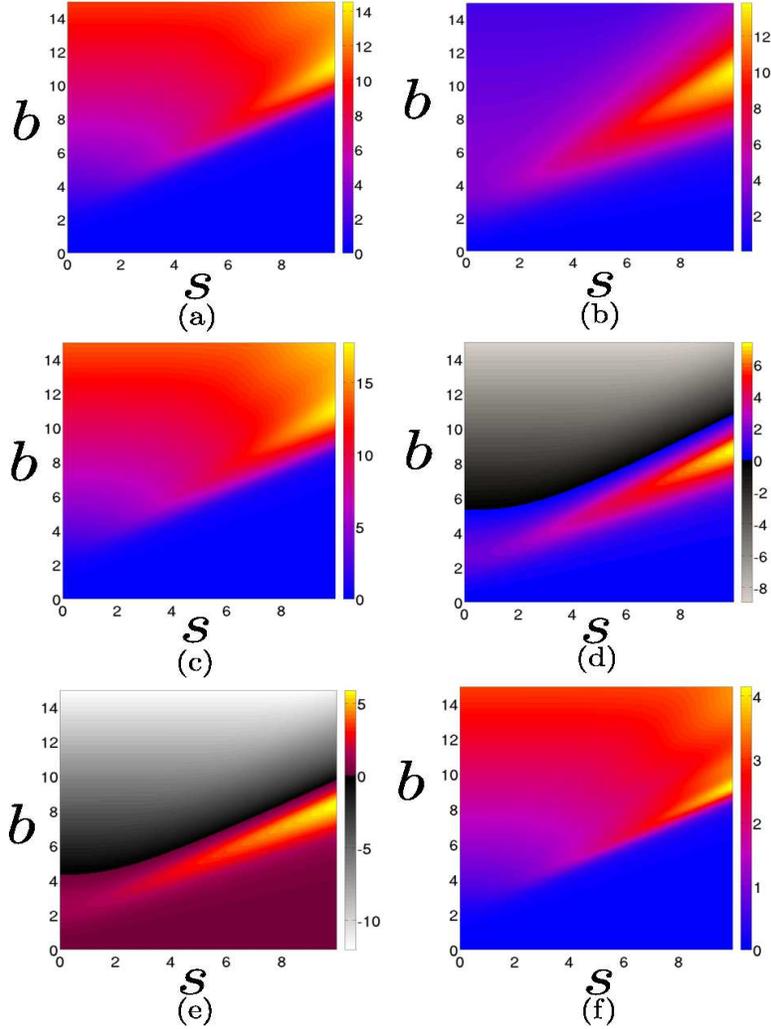}
\caption{\textbf{Highly sensitive tumors ($\bf{\rho=1.0~\text{mg}^{-1}}$)}. The fraction of cells that have survived the treatment $\sigma_{i}=P(t_{\text{nad}})/P(0)$ at the time $t_{\text{nad}}$ when the tumor reaches its minimum size (nadir) is computed for each protocol $\text{p}_{i}$. Then the same computation is repeated for a $300 \times 300$ square grid of values in the parameter set $(s,b)$. Finally, the relative destruction between two protocols $\text{p}_{i}$ and $\text{p}_{j}$ is represented in the parameter space by computing the natural logarithm (for which the results are the most clear) of the quotient $\sigma_{i}/\sigma_{j}$ (color bar). If $\log(\sigma_{i}/\sigma_{j})>0$ in the color bar, the protocol $\text{p}_{j}$ is better than protocol $\text{p}_{i}$. (a) The effects of increasing the dose represented by $\sigma_{1}/\sigma_{2}$, which is always beneficial. (b) The case $\sigma_{1}/\sigma_{3}$ representing an increase in dose-density, which is beneficial for all parameter values. (c) Longer continuous infusions are studied by computing $\sigma_{1}/\sigma_{4}$, which is the most beneficial. (d) Dose-intensity versus dose-density is studied by computing $\sigma_{2}/\sigma_{3}$. Depending on the parameter values one strategy is better than the other. (e) The values of $\sigma_{4}/\sigma_{3}$, comparing an increase in dose-intensity against an increase in the time of infusion. (f) The case $\sigma_{2}/\sigma_{4}$ shows that a sufficient increase in the duration of infusion is better that increasing the dose-intensity.}
\label{fig:6}
\end{figure}

Obviously, when the sensitivity to the drugs is high, we see that increasing the dose delivered or the frequency of the cycles three times is beneficial for all parameter values. This is shown in Fig.~\ref{fig:6}, where the survival fraction at the nadir of the treatment is represented in the parameter space. Thus, mathematically, increasing the frequency of the cycles and the dose of drug are both convenient. However, as can be seen in Fig.~\ref{fig:6}(b), if the drugs are very effective destroying the tumor cells (high values of $b$), the benefits of increasing dose-density (the frequency) are very small ($\sigma_{1}/\sigma_{3}$ close to one). The reason that explains this behavior is that if the cell population is severely reduced, then the regrowth of the tumor cells is comparatively small. This does not occur for an increase in the dose-intensity (see Fig.~\ref{fig:6}(a)), although increasing too much the dose might introduce an intolerable toxicity. Consequently, the relative benefits of increasing three times the dose, compared to an identical increase in the frequency of administration, are higher for small values of $s$ and high values of $b$. This is shown in Fig.~\ref{fig:6}(d), where we clearly see that $\sigma_{2}$ is smaller than $\sigma_{3}$ for such values of $b$. 

We also see in the same figure that as we increase $s$, dose-densification becomes more pertinent. Indeed, the Norton-Simon hypothesis strengthens the importance of dose-density. Ultimately, an increase in the duration of the infusion to half a week, maintaining constant the dose and the periodicity, seems to be the most beneficial in our model (see Figs.~\ref{fig:6}(e) and (f)). In part this occurs because the drug is acting more time on the tumor, avoiding the regrowth between cycles. However, it must be noted that the peak of drug concentration is smaller. Therefore, it is expected that this is true only for tumors which are very sensitive to the drugs and when these drugs are administered at sufficiently high doses. Furthermore, this is somewhat equivalent to deliver drugs continuously at doses that are smaller than the standard, which in many situations can be more toxic and ineffective \cite{dan}.

\subsection{Low sensitivity scenario}\label{sec:lrs}

\begin{figure}
\centering
\includegraphics[width=0.8\linewidth,height=0.75\linewidth]{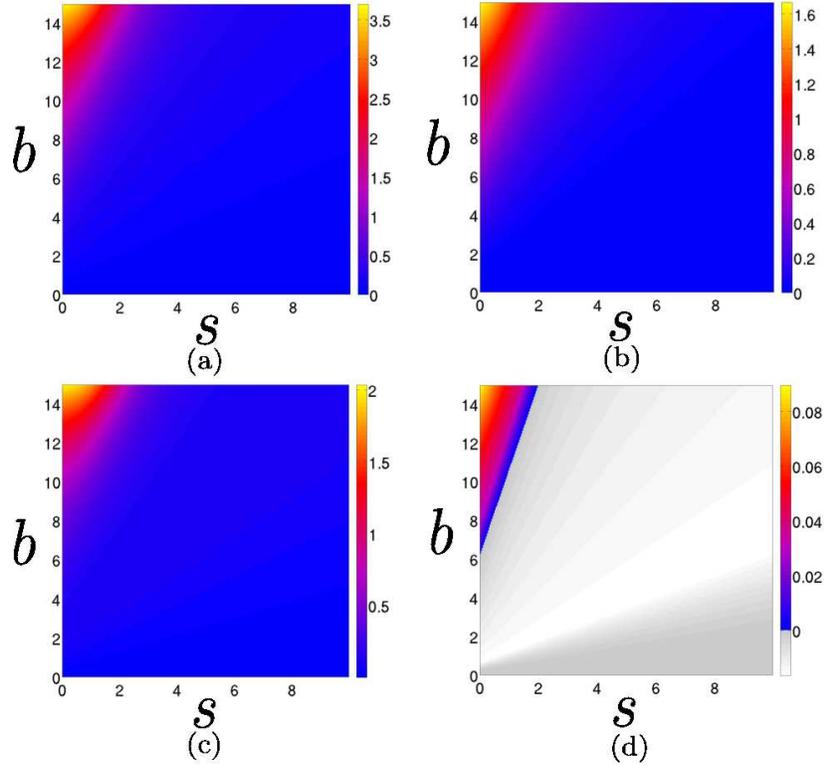}
\caption{\textbf{Slightly sensitive tumors ($\bf{\rho=0.01~\text{mg}^{-1}}$)}. Again, the fraction of cells that have survived the treatment $\sigma_{i}=P(t_{\text{nad}})/P(0)$ at the time $t_{\text{nad}}$ when the tumor reaches its minimum size (nadir) is computed for each protocol $\text{p}_{i}$. Then the same computation is repeated for a $300 \times 300$ square grid of values in the parameter set $(s,b)$. Finally, the relative destruction between two protocols $\text{p}_{i}$ and $\text{p}_{j}$ is represented in the parameter space by computing the natural logarithm of the quotient $\sigma_{i}/\sigma_{j}$ (color bar). If $\log(\sigma_{i}/\sigma_{j})>0$ in the color bar, the protocol $\text{p}_{j}$ is better than protocol $\text{p}_{i}$. (a) The effects of increasing the dose represented by $\sigma_{1}/\sigma_{2}$ are always beneficial. (b) The case $\sigma_{1}/\sigma_{3}$ representing an increase in dose-density, which are barely beneficial. (c) Dose-intensity versus dose-density is studied by computing $\sigma_{3}/\sigma_{2}$. As can be seen, dose intensification is always a better strategy in this scenario. (d) Longer continuous infusions $\sigma_{1}/\sigma_{4}$ represents now a very bad strategy.}
\label{fig:7}
\end{figure}

Now we turn our attention to a situation in which the sensitivity of the tumor cells to the drugs is low. In this case the scenario changes dramatically and becomes less rich. Firstly, in Fig.~\ref{fig:7} we observe that the effects of increasing the dose of drug or the frequency are in general much smaller. More importantly, increasing the dose three times is now more convenient than increasing dose-density by the same amount for all the values of the parameters $b$ and $s$. As can be seen in Fig.~\ref{fig:7}(c), $\sigma_{3}>\sigma_{2}$ everywhere in the parameter space. This result can be interpreted as follows. If the sensitivity of the tumor cells to the drugs is very low, the more dose of drug we give, the more chances for a tumor cell to be lethally hit by these drugs. Consequently, if the drugs are not causing substantial destruction, increasing the frequency does not introduce significant advantages. For the same reason, we see in Fig.~\ref{fig:7}(d) that increasing the time of drug infusion is now harmful, except for drugs that are very destructive in the absence of the Norton-Simon effect.

\subsection{Testing the robustness of our results}\label{sec:lrs}

We have tested the effect of modifying the parameters that represent the carrying capacity $K$ and the rate of elimination of the drug $k$. Increasing the carrying capacity one thousand times (which approximately corresponds to a tumor of one kilogram) does not alter substantially the conclusions. Now the growth of the tumor approximates to exponential and the model is exactly the same as the one presented in \cite{gard}. Mathematically, this can be written as $\gamma(P) \rightarrow r$ and $h(P) \rightarrow r b$. Consequently, the parameter $s$ loses its importance and the graphics are similar to the ones appearing in Fig.~\ref{fig:6} for small values of $s$. Increasing four times the value of $k$, which reduces the half-life to six hours, introduces two main effects. Firstly, the drugs spend less time in the organism and therefore the impact of chemotherapy is reduced. This enhances the importance of dose-density. The convenience of increasing the frequency over increasing the dose for highly sensitive tumors is more noticeable. Secondly, and for the same reason, the benefits of increasing the time of infusion are higher for the high sensitivity case. 

\subsection{Other functional forms}
\label{sec:lrs2}
To conclude our study, we have also tested the robustness of our results by modifying the functional form of the growth term $\gamma(P)$. If we consider the Gompertz differential equation, then we have $\gamma(P)=r \log(K/P)$. We keep the same value of the carrying capacity $K=10^{9}~\text{cells}$ but modify the parameter $r=0.12~\text{weeks}^{-1}$. This value is chosen so that two tumors, one following Gompertzian growth and another growing logistically, reach a size of half of its maximum value $K$ at the same time. 

\begin{figure}
\centering
\includegraphics[width=1.0\linewidth,height=0.6\linewidth]{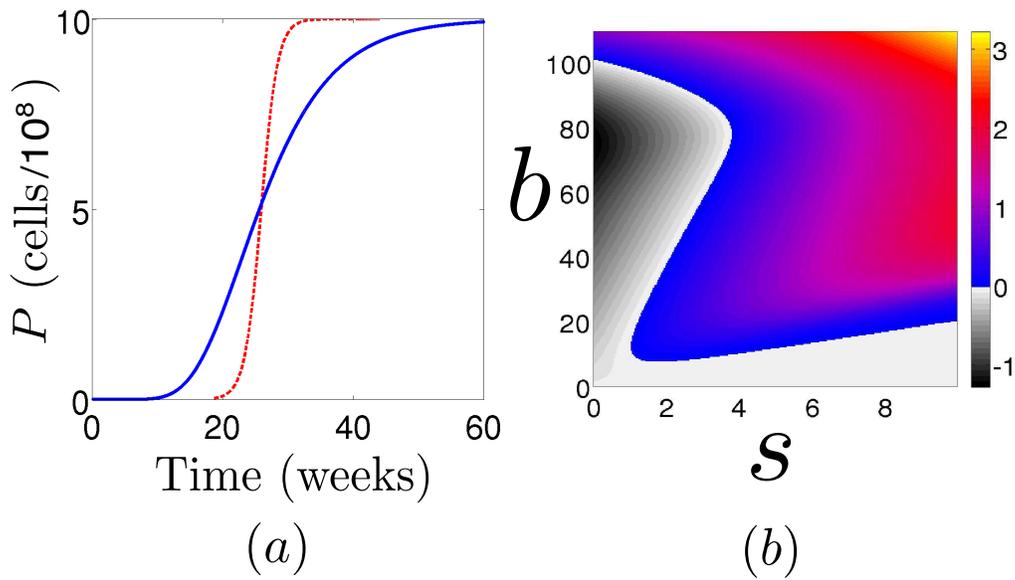}
\caption{\textbf{Logistic vs Gompertz}. (a) A comparison between a logistic growth (red) given by the equation $\gamma(P)=0.8~P (1-P/K)$ and a Gompertzian growth (blue) given by $\gamma(P)=0.12~P \log(K/P)$. The initial condition is $P_{0}=1~\text{cell}$, for both time series. The parameters have been adjusted so that the time at which $P=K/2$ is the same. As can be seen, the logistic growth is more steep. (b) The dose-intense protocol $\text{p}_{2}$ is compared to the dose-dense protocol $\text{p}_{3}$ by computing $\sigma_{3}/\sigma_{2}$ for the Gompertz type of growth and following exactly the methods explained in the two previous figures. The color bar represents $\log(\sigma_{3}/\sigma_{2})$, and it is positive when the protocol $\text{p}_{2}$ is more effective than the protocol $\text{p}_{3}$.}
\label{fig:8}
\end{figure}

As can be seen in Fig.~\ref{fig:8}(a), under these conditions, the Gompertzian growth is less steep than the logistic. If we compare $\dot{P}$ for the two types of growth and the parameters given, we see that the regrowth of the tumors between cycles is smaller for the Gompertz case, except for very small cell populations $P$. Consequently, and as depicted in Fig.~\ref{fig:8}(b), the benefits of trebling the dose-density relative to augmenting the dose three times are in general smaller. Nevertheless, we insist that increasing the dose and the frequency are still advantageous for both scenarios. It is remarkable that for small values of $s$, dose-density prevails over dose-intensity. As previously said, this is a consequence of the fact that for small values of the cell populations and the parameter values used in the simulation, the regrowth is greater for the Gompertz case. Anyway, this should not be read as a failure of the Norton-Simon hypothesis, since the same stands for sufficiently high values of $s$.

Finally, we have also inspected the effects of modifying $h(P)$, which we recall that models the Norton-Simon effect. In particular, we have considered the case $h(P)=r(1-s P/K)$. For $s=1$, we obtain a direct proportionality between $h(P)$ and $\gamma(P)$, which represents the most rigorous form of the effect. No substantial changes have been found in this case, except a less sensitivity to the parameter $s$, similar to the one described two paragraphs above.

\section{Concluding Remarks}\label{sec:con}

In the present work we have studied the relative benefits of dose-dense protocols and the role of the Norton-Simon hypothesis in nonlinear cancer chemotherapy. For this purpose, we have devised the simplest mathematical model capable of representing these protocols in a clear manner. More sophisticated multicompartment models including the evolution of resistance, combination therapy, toxic side-effects and stochastic fluctuations can be developed in a straightforward way using the present model as a starting point.

Our results demonstrate that dose-dense protocols should be generally beneficial, specially if the Norton-Simon hypothesis stands. The main idea underlying this hypothesis is that chemotherapy is less effective on slowly growing tumors. In particular, if the rate of tumor growth decreases with tumor size, the rate of growth just before a cycle of chemotherapy starts is lower than the rate of growth right after it, and therefore regrowth might dominate, leading to unsuccessful treatment. However, when most of the cancer cells are intrinsically insensitive to the drugs, our results suggest that dose-density might be barely useful, compared to dose escalation. As a rule of thumb, it can be said that dose-density should be more helpful when cancer cells are very chemosensitive and the regrowth between cycles is considerable, relative to the destruction caused by the cytotoxic drugs.

It is notable that our investigation insinuates that the unresponsiveness of a tumor to chemotherapeutic drugs can be broadly classified into two categories. The first could be termed intrinsic sensitivity, and depends on all types of biomolecular mechanisms that a single tumor cell possess to circumvent the destructive effects of a cytotoxic drug. In our model, this sensitivity is represented by the parameter $\rho$, and leads to the concept of effective dose $\rho D$, which was introduced in previous works \cite{ddpc}. The second type of sensitivity could be termed extrinsic sensitivity, which emerges as a consequence of how a tumor organizes as a whole and its morphology. We believe that it is this second class of sensitivity that should be linked to the Norton-Simon hypothesis and, consequently, the parameter $s$ would model an unresponsiveness of this nature.

To conclude, the results for highly sensitive tumors also point to the fact that the optimal situation should be obtained when the regrowth between cycles is avoided by increasing the duration of the infusions and administering drugs in an almost continuous fashion. Since continuous protocols can be too toxic and lead to tachyphylaxis, the present work indicates that perhaps other means of avoiding regrowth between cycles should be devised, instead of increasing the frequency of the protocol. For example, a good strategy could be to render the tumor cells quiescent after chemotherapeutic drugs have caused their damage. This could be achieved by means of some targeted cytostatic drug \cite{ddpc}. Then, hours after the interruption of the cytostatic effects \cite{rp}, the cells would abandon the $G_{0}$ phase and reenter the cell cycle, to suffer another destructive wave of cytotoxic chemotherapy.

\section*{Aknowledgments} \label{sec:aknow}

This work has been supported by the Spanish Ministry of Economy and Competitiveness and by the Spanish State Research Agency (AEI) and the European Regional Development Fund (FEDER) under Project No. FIS2016-76883-P. K.C.I. acknowledges FAPESP (2015/07311-7 and 2017/20920-8).

\bibliographystyle{plain}
\biboptions{numbers,sort&compress}

\end{document}